\documentclass[prd,twocolumn,floatfix,showpacs,showkeys]{revtex4-1}
\usepackage{latexsym,amssymb,amsmath,amsfonts}
\usepackage[mathscr]{euscript}
\usepackage{graphicx}
\usepackage{bm}
\usepackage{multirow}
\usepackage{setspace}
\newcommand{\beq}{\begin{equation}}
\newcommand{\eeq}{\end{equation}}
\newcommand{\bc}{\begin{center}}
\newcommand{\ec}{\end{center}}
\newcommand{\eeqa}{\end{eqnarray}}
\newcommand{\beqa}{\begin{eqnarray}}

\newcommand{\ra}{\rightarrow}

\newcommand{\al}{\alpha}
\newcommand{\be}{\beta}
\newcommand{\ga}{\gamma}

\newcommand{\de}{\delta}

\newcommand{\ep}{\epsilon}

\newcommand{\la}{\lambda}

\newcommand{\si}{\sigma}

\newcommand{\ph}{\phi}

\newcommand{\ed}{\end{document} }

\begin{document}

\title{Gravitational dipole moment from an exact solution with torsion}
\author{Richard T. Hammond}
\email{rthammond@ncat.edu }
\affiliation{
North Carolina A\&T State University and\\
University of North Carolina at Chapel Hill\\
Chapel Hill, North Carolina\\
Greensboro, North Carolina}

\date{\today}

\begin{abstract}
Dipole fields are common in electromagnetism and may be viewed as the result of a positive and negative charge (or pole) which are close together.  A dipole field in gravity is not expected to exist because negative mass has never been observed. However,
an exact solution to the gravitational field equations with torsion is presented  that does correspond to a dipole gravitational field, even though there is no negative mass.  The theory is based on a non-symmetric metric tensor, and it is found there is a singularity in the metric tensor at the origin but no event horizon. Gravitational dipoles have been used to solve the dark matter problem. 
\end{abstract}
\maketitle

\section{Introduction}

The surprising result, a gravitational dipole field, is the result of a theory of gravity with a non-symmetric tensor and a stringy source.
Einstein devoted many years of his life working on gravitation with a non-symmetric metric tensor (NMT)\cite{einothers}. Writing the metric tensor in terms of the symmetric and antisymmetric parts, $g_{\mu\nu}=\ga_{\mu \nu}+\ph_{\mu\nu}$ respectively,  he tried to associate $\ph_{\mu\nu}$ with electromagnetism. Today this is generally regarded as a failure and, even though Moffat\cite{mof} developed a theory using NMT without electromagnetism, the  notion of NMT in general has acquired an aura of failure. In fact, Pauli stated, ``it is a completely sucked out lemon."

Perhaps the biggest problem with the theories above is they lacked ``proper'' physical motivation. Einstein believed he  found one in his Hermitian theory.\cite{ein} The physical significance of this, which allowed a reduction in the number of potential terms, was charge conjugation invariance. This certainly sounds like a physical motivation, but it was not, because it already assumes the antisymmetric part of the metric tensor has something to due with electromagnetism.

``Proper'' physical motivation came from 
Papapetrou who, following the NMT, reasoned if the metric tensor were not symmetric, then the energy momentum tensor could not be symmetric.\cite{pap},\cite{bel} He showed the antisymmetric part of the energy momentum tensor must be due to intrinsic spin.

Now we have motivation for the NMT. Since particles have spin in addition to mass, one is naturally led to the notion that the metric tensor is non-symmetric. (Of course this is for particles, since for  macroscopic objects the spins can cancel).
A few years later Sciama considered the notion of spin from a NMT, but soon after abandoned the idea and considered torsion (the antisymmetric part of the affine connection) with a symmetric metric tensor to describe spin.\cite{sci}
With the antisymmetric part of the NMT being related to spin, and since the antisymmetric part of the NMT gives rise to torsion, we should expect torsion is related to spin.

 This was already believed to be the case even with a symmetric metric tensor,\cite{heh} and  was proven  by showing the only way to achieve conservation of total angular momentum was with torsion arising from intrinsic spin.\cite{stf} Gravitation with a symmetric metric tensor and non-symmetric affine connection are referred to as metric-affine theories.\cite{nmt} However,  these were sometimes charged as being ad hoc, in that the affine connection does not have to be non-symmetric if the metric tensor is symmetric. These charges are dismissed with the NMT and, in fact, it may be argued in reverse,  that a symmetric tensor may be ad hoc. 
For reviews of torsion see,\cite{heh},\cite{hamrev}\cite{shap} for reviews of the NMT see \cite{nmt}.

Today we have a much different perch from which to view NMT theory. With the Standard Model firmly in place, we can step away from the rusty conviction that the antisymmetric part of the NMT is related to electromagnetism. In addition, we have new insight from string theory. As shown below, what many early researchers thought was a connection between NMT and electromagnetism is really a connection between NMT and spin, or the Kalb-Ramond field, as described below.

Recently a theory was published using these ideas in which the antisymmetric part of the NMT was the potential for torsion.\cite{ham2019} In this case, the torsion,  $S_{\al\be\sigma}$, is given by $S_{\al\be\sigma}=
\ph_{\al\be,\sigma}+\ph_{\sigma\al,\be}+\ph_{\be\sigma,\al}$. Thus, the NMT theory turns out to be a theory of gravity and spin, not a theory of gravity and electromagnetism.  It was shown the NMT theory with spin in \cite{ham2019}, in the weak field limit, reduced to the theory with a symmetric metric tensor torsion of the string theory type--Kalb Ramond field, as given above.\cite{kalb}, \cite{scher}

However, in the metric-affine theories the torsion is gauge invariant, i.e., invariant under the transformation $\ph_{\mu\nu}\ra  \xi_{\mu,\nu}-\xi_{\nu,\mu}$.  When the potential is the antisymmetric part of the metric tensor this invariance is lost. The same thing happens in string theory with open strings. In order to maintain gauge invariance electric charge is placed at the ends of the string and it is assumed under the gauge transformation given above, the electromagnetic potential transforms as $A_{\mu}\ra A_{\mu}+\xi_\mu$.

Thus, we end up with a theory of gravitation, spin, and electromagnetism, although this is not the unified theory Einstein sought. More details may be found in Ref. \cite{ham2019}.
It was also shown in the weak field limit this theory reduces to the metric-affine case in which the metric tensor is symmetric and the torsion is given by the form given above.

Einstein believed, or hoped, the NMT would provide a doorway to quantum phenomena, or at least that there was a connection between the NMT and quantum mechanics. In this view he was right. This is because spin had always been thought of as purely quantum in nature, but here it takes on a classical description.\cite{cla}

The point of this paper is to present an exact solution to a special case of the  NMT of \cite{ham2019} in the weak field limit. The electromagnetic field is ignored and the cosmological constant is assumed to vanish. As stated above, in the weak field limit the NMT theory reduces to metric-affine work, which will be used here.\cite{hamrev}
To proceed, first note that with a totally antisymmetric torsion, an additional term to the Lagrangian is often considered,
a term given by $\Lambda  S_{\al\be\sigma}S^{\al\be \sigma}$,  $\Lambda$ is a constant, so 
the variational principle for the more general Lagrangian density is given by

\beq
\de\int d^4x\sqrt{-g}(R+\Lambda  S_{\al\be\sigma}S^{\al\be \sigma})=0
\eeq
where $R$ is the curvature scalar of $U4$ space-time. The case with sources has been studied extensively\cite{hamrev} and it was shown strings are the most natural source,\cite{hamstr} but here we consider only the vacuum solution.

 It is useful to separate the equations into the Riemannian  part plus torsion, which yields,

\beq\label{gfe}
G^{\mu\nu}=\la t^{\mu\nu}
\eeq
where $\lambda=\Lambda -1$ and
\beq\label{tfe}
S^{\mu \nu \sigma}_{\ \ \ \ ;\sigma}=0
\eeq
where the semicolon represents the Levi-Civita covariant derivative
and where $G^{\mu\nu}$, and $R_{\mu\nu}$ below, are the Riemannian Einstein and Ricci tensors, and

\beq
t^{\mu\nu}=\frac12 g^{\mu\nu}
S^{\al\be\si}S_{\al\be\si}
-3S^{\mu\al\be}S^\nu_{\ \al\be}
.\eeq

At this point we can think of $t^{\mu\nu}$ as the energy  momentum tensor in a Riemannian space.
We assume the metric tensor and torsion are functions of $r,\theta$ only, and using

\beq
b_\mu \equiv \ep_{\mu\al\be\ga}S^{\al\be\ga}
\eeq
where $\ep_{\mu\al\be\ga}$ is the totally antisymmetric tensor, (\ref{tfe}) becomes

\beq\label{bfe}
b_{\mu;\nu}-b_{\nu;\mu}=0
.\eeq

We are seeking a solution in coordinates exterior to a static intrinsic spin. It has been emphasized nothing is rotating\cite{stf} so we may assume $g_{0n}=0$ ($n=1,2,3$), and  assume the metric tensor takes the form

\beq
ds^2=A(r,\theta)dt^2-B(r,\theta)dr^2-C(r,\theta)(r^2d\theta^2+r^2\sin^2\theta d\ph^2)
.\eeq
With this  we see (\ref{bfe})  may be solved exactly. The covariant tensor components are

\beq\label{bsol}
b_1=\frac{2K}{r^3}\cos\theta,\ \ \ \ \ b_2=\frac{K}{r^2}\sin\theta
\eeq
where $K=3kS/4\pi$ where $S$ is the taken to be the spin of a spin 1/2 particle and $k=8\pi G$.

With this, the field equations (\ref{gfe}) become

\beq\label{gfe2}
R_{\mu\nu}=\frac{-\la}{6} b_\mu b_\nu
,\eeq
where $R_{\mu\nu}$ is the Ricci tensor of Riemannian geometry. This becomes

\beq
R_{11}=\frac{-2\la K^2 \cos^2\theta}{3r^6}
\eeq

\beq
R_{22}=\frac{-\la K^2 \sin^2\theta}{6r^4}
\eeq

\beq
R_{12}=\frac{-\la K^2 \sin^2 2\theta}{6r^5}
\eeq
with the rest of the $R_{\mu\nu}=0$, and the solution to (\ref{gfe2}) is

\begin{widetext}
\beq\label{gsol}
g_{\mu\nu}=
\left(
\begin{array}{cccc}
 e^{q\cos\theta/r^2} &  0 &  0 & 0\\
  0&  -e^{-q\cos\theta/r^2}, &  0 &0 \\
 0 &  0 & -r^2e^{-q\cos\theta/r^2}  & 0\\
 0 & 0& 0&-r^2\sin^2\theta e^{-q\cos\theta/r^2}\\
\end{array}
\right)
\eeq
\end{widetext}
where $q=\la K^2/3$ and the first entry is $g_{00}$, and $g_{11}$ is displayed in Fig.1.

\begin{figure}
\includegraphics[width=0.4\textwidth]{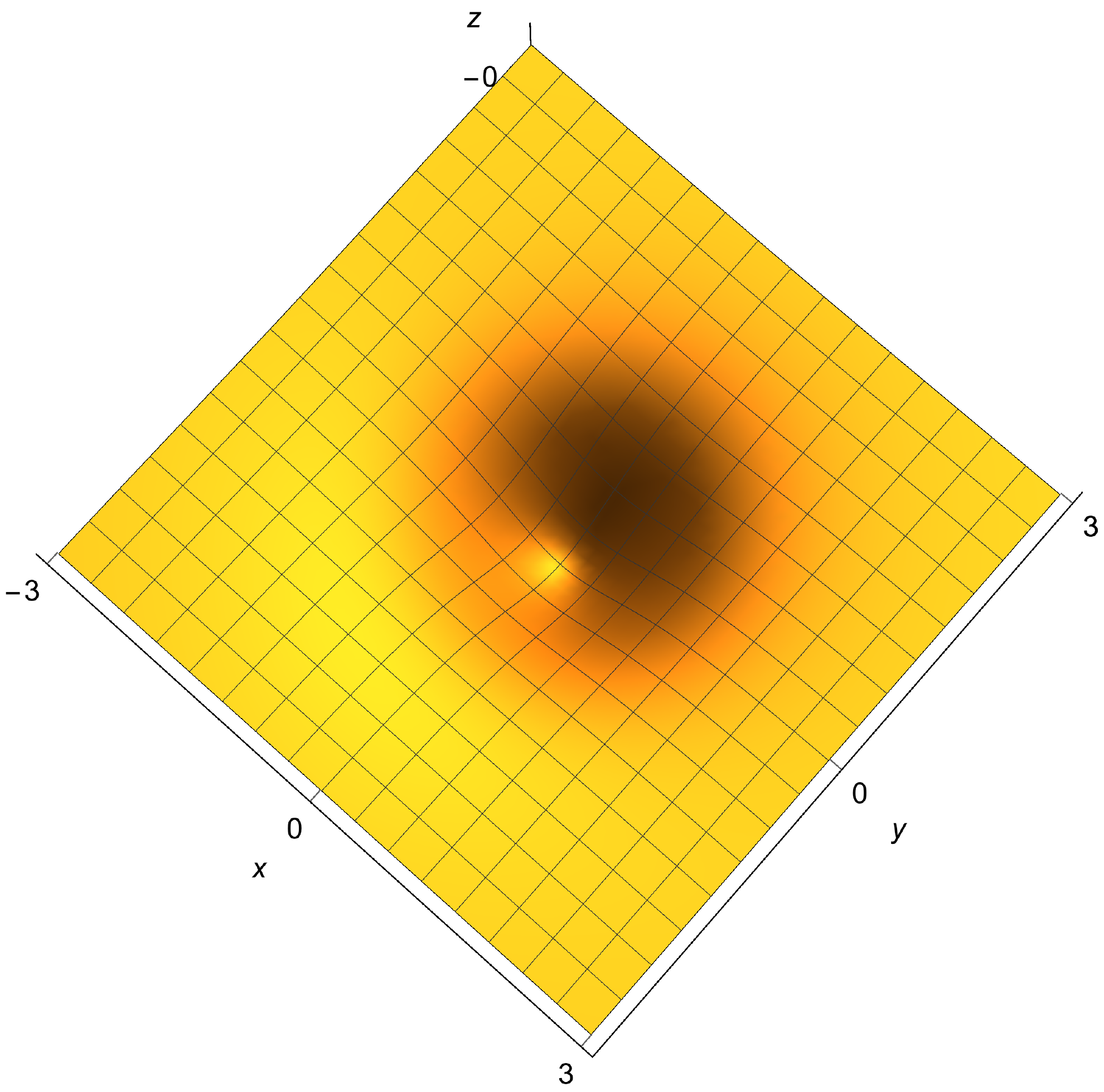}
\caption{In Cartesian coordinates, plot of $g_{11}$ for constant $z$ near  $0$ with $q=1$.}\label{mt}
\end{figure}

This solution has no event horizon and is Minkowskian in the far field. It is singular at the origin so we have a naked singularity. It should also be noted that this does not reduce to a Schwarzschild solution as the spin goes to zero so these are truly dipole particles.
There are two killing vectors, which are

\beq
\xi_t=  e^\frac{q\cos\theta}{r^2} \ \ \ \ \xi_\ph=e^\frac{-q\cos\theta}{r^2} \ \ \ \
.\eeq

Since this a stationary (in fact, static) solution,  the Landau Lifshitz\cite{ll} formula may be used, which for low velocity is

\beq\label{force}
{\bm f} =-mc^2{\bm \nabla}\ln\sqrt{g_{00}}
,\eeq
which gives the force on a particle of mass $m$ in the field of (\ref{gsol}). A potential, $\ph$, may be defined so that, calling 
${\bm f} =-m{\bm \nabla}\ph$
it is found

\beq\label{pot}
\ph=\mu_g
\frac{\cos\theta}{r^2}
\eeq
where $\mu_g= q$ is the gravitational dipole moment.

The field of a dipole is well-known, but it comes as a surprise in general relativity without negative mass. Even in this case, the center of mass can be redefined so there does not seem to be a way for GR to account for a gravitational dipole. However, with torsion we see how this arises. Experimental searches have been conducted but no evidence of a dipole moment survives scrutiny.\cite{saf}  However, these searches really look for a $1/r^2$ component of the potential energy as a perturbation. The magnetic dipole has no monopole term and neither does the gravitational dipole described above.
In gravitation, the strangest thing about a dipole field is that there are  regions of attraction {\it and} repulsion, as (\ref{force}) and (\ref{pot}) show.

It has been shown a galaxy full of gravitational dipoles can solve the dark matter problem.\cite{bl1},\cite{bt1},\cite{bt2} The dipole distribution acts like Milgrom's modified gravity theory and explains the flat rotation curves of galaxies. In that work the authors were forced to postulate the existence of dipole particles, but here they arise naturally.
 
In the above, it was assumed the dipole field is related to an elementary particle, hence the spin was taken to that of a spin $\hbar/2$ particle. But this could also represent the field far from a cosmic string, in which case the constant $K$ is unknown, and could be very large. To see this, note that the assumption leading the result (\ref{bsol}) was the symmetry. In the far field, a loop in a plane has the same symmetry of a dipole.

In this case, the cosmic string with torsion could give rise to regions with repulsive gravity. In addition, such cosmic strings could be natural particle accelerators. To see this, the Newtonian case  may be considered. Suppose a particle is created a distance $a$ from the galactic plane, and is repelled to infinity without collisions (an ideal case). Then, from (\ref{pot}), we see the change in the potential is equal to $\mu_g/a^2$ (where $\mu_g$ is  the dipole moment of the cosmic string).

In summary, an exact, non-spherically symmetric  solution to the field equations with torsion has been presented. The solution was shown to give rise to a dipole field for torsion, and also a dipole gravitational field. It was noted there is a singularity but no event horizon, it was speculated cosmic strings with torsion could give rise to repulsive gravity in galaxies, and that such strings could act as cosmic particle accelerators.

\ed